\documentstyle[pra,aps,preprint]{revtex}
\begin{document}
\draft
\title{From Tomonaga-Luttinger to Fermi liquid 
in transport through a tunneling barrier. }
\author{V.V. Ponomarenko$^{1,2}$ and N. Nagaosa$^{1}$}
\address{$^1$ Department of Applied Physics, University of Tokyo,
Bunkyo-ku, Tokyo 113, Japan\\
$^2$ A.F.Ioffe Physical Technical Institute,
194021, St. Petersburg, Russia}
\date{\today}
\maketitle
\begin{abstract}
Finite length of a one channel wire results in crossover from 
a Tomonaga-Luttinger to Fermi liquid behavior with lowering
energy scale. In condition that voltage drop $(V)$
mostly occurs across a tunnel barrier inside the wire we found
coefficients of temperature/voltage expansion of low energy
conductance as a function of constant of interaction, right and
left traversal times. At higher voltage the finite length 
contribution exhibits oscillations related to both traversal times
and becomes a slowly decaying correction to the scale-invariant 
$V^{1/g-1}$ dependence of the conductance.

\end{abstract}
\pacs{72.10.Bg, 72.15.-v, 73.20.Dx}

\narrowtext
Quantum transport in Tomonaga-Luttinger liquids (TLL) has
attracted a great deal of interest as it was 
suggested to be realized in a 1D constriction 
\cite{ar,kf} 
and the edge state of the fractional quantum Hall (FQH) 
liquid \cite{fqh}. Both suggestions have been supported in
recent experiments on FQH \cite{hallex} and on 1D \cite{tar} transport.
Even more experimentalists \cite{1dex} made claim on observation of the
interaction effects in the 1D transport, however, without comprehensible
connection with a theoretical model. Unless there is a resonant tunneling
the repulsive interaction typically suppresses conductance at low energy.
The experiment by Tarucha's group \cite{tar} on 1D transport through a long
wire with a weak impurity random potential inside demonstrated a crossover
from TL liquid to Fermi liquid behavior at low temperature. 
This crossover is a finite length effect \cite{of} and may be 
described in an inhomogeneous TLL model (ITLL) \cite{mas,pon,safi}. The 
ITLL model predicts both the conductance behavior  in
the Fermi liquid region up to a renormalization constant
(i. e., relations between the coefficients at  
temperature/voltage in different integer degrees)  and the interaction 
dependent non-analytical  behavior in the TLL region.

The aim of this work is to examine this crossover in
the low voltage conductance of the 
one channel wire where its suppression is mostly determined by a high point 
barrier located inside the wire. Position and height of the barrier are 
assumed to be due to an external gate. Therefore both distances $L_{R(L)}$
from the barrier to the right/left reservoir and the ratio 
$\zeta=L_R/(L_R+L_L)$ are 
assumed to be known. Then it will be shown below that just two more parameters,
which are the total traversal time $t_L$ equal to sum of the right and left 
ones:$t_0=t_{R}+t_{L}$ and the constant of the forward scattering
$g$, are necessary for desciption of the crossover, and that the ITTL model gives
quite a few ways to determine these parameters 
from either high voltage $(V>1/t_0)$ or
low voltage $(V<1/t_0)$ conductance measurements.
Temperature dependence of the conductance 
in this model was considered in \cite{fn}.
Similar model turned out to be useful for experimental study of
the FQH transport \cite{hallex}. It was noticed recently \cite{chkl}
that  there is a special way of connection between a $\nu=1/3$ FQH liquid
and leads when the whole setup 
corresponds to the $g=1/3$ ITLL model.
If so, the results obtained below 
for spinless electrons could be directly addressed to that FQH device
when the FQH liquid embeds a point scatterer.

Under condition that the link between
two parts of the wire is weak enough it suffices to 
apply the tunneling Hamiltonian approach in the lowest order
to describe the transport \cite{kf,gui,es}.
Current flowing 
through the weak link located, say, at $x=0$ inside the wire 
is given by the operator 
$J(t)=-i [A \psi^+_R(0,t) \psi_L(0,t)-h.c.]$, $(e,\hbar=1)$,
 where $A$ is the tunneling amplitude
and $\psi_{R,L}(x,t)$ are the electron annihilation operators in 
the right $0 \leq x <L_R$ and in the left $-L_L<x \leq 0$ part of the 
wire,respectively. The average current 
under voltage $V$ applied to the left lead can be written as:
$<J>=2 \pi |A|^2 \int d \epsilon [f(\epsilon-V)-f(\epsilon)] \rho_R(\epsilon)
\rho_L(\epsilon-V)$, 
  where $f$ is Fermi distribution. The problem reduces to finding of
the tunneling density of states
of the right (left) end of the junction $\rho_{R(L)}(\epsilon)$ which are
the sum of the particle $\rho_{p}(\epsilon)=
(1-f(\epsilon)) \rho(\epsilon)$ and hole $\rho_{h}(\epsilon)=
f(\epsilon) \rho(\epsilon)$ densities. The latters relate to
the particle correlator as 
$\rho_p(\epsilon)=1/(2 \pi) \int d t e^{i \epsilon t}
<\psi(0,t) \psi^+(0,0)>$ and to the hole one as 
$\rho_h(\epsilon)=1/(2 \pi) \int d t e^{i \epsilon t}
<\psi^+(0,0) \psi(0,t)> $. 

{\it Tunneling density of states} -  
To calculate the tunneling density of states  $\rho_{R}$ 
on the right side of the weak link
let us first consider spinless fermions and apply
bosonization to the $\psi$ field under condition of an elastic reflection 
from the boundary located at $x=0$ \cite{f,o}. 
( Carrying out this calculations
we will omit index "R" below. )
Bosonic repersentation  of the $\psi$ field reads 
$\psi(x,t)=\sum_{a=r,l} \psi_{a}(x,t)=(2 \pi \alpha )^{-1} 
\sum_\pm exp\{i(\theta(x,t) \pm \phi(x,t))/2 \}$, where $\psi_{r(l)}$ is
the right (left) going chiral component of $\psi$ and the 
$\theta$ and $\phi$ fields are bosonic and mutually conjugated 
$[\theta(x,t),\phi(y,t)]=2 \pi i sgn(x-y)$. 
The elastic reflection means that $\psi_{l}(0,t)= e^{i\delta }\psi_{r}(0,t)$
with an appropiate phase shift $\delta $. This results in both:
\begin{equation}
\phi(0,t)=\delta, \;\; \frac{1}{2\pi} \partial_x \theta(x,t)|_{x=0}=
\psi^+_r(0,t)\psi_{r}(0,t)-\psi^+_{l}(0,t)\psi_{l}(0,t)=0.
\label{1}
\end{equation}
Then the density of particle states could be found as
\begin{equation}
\rho_p(\epsilon)=\frac{\rho_O E_F}{2 \pi} 
 \int^{+\infty}_{-\infty} d t e^{i \epsilon t +\frac{1}{4}
[<\theta(0,t)\theta(0,0)>-<\theta^2(0,0)>]}
\label{2}
\end{equation}
where the value of the free electron tuneling density was 
introduced as: $\rho_O=(1+\cos \delta )/(\pi v)$.

The problem reduces to finding the $\theta$ field correlator. It 
can be done for the finite length piece of the wire adiabatically 
connected to the lead making use of the ITTL model \cite{mas,pon,safi}. 
In this model
the Tomonaga-Luttinger interaction $(\sum_{r,l}\rho_a)^2$ is switched
on in the wire $x<L_R$ and switched off outside. Then the Hamiltonian
takes a bosonized form
\begin{equation}
{\cal H}= \int_0^{\infty} dx  \frac{v}{2}
\{u^2(x)  
\left({{\partial_x \phi(x) } \over {\sqrt{4 \pi}}} \right)^2 + 
\left({{\partial_x \theta(x) } \over {\sqrt{4 \pi}}}
\right)^2 \}
\label{3}
\end{equation}
where function $u(x)$ ensuing from the interaction can be approximated
in the low energy limit by a step-function: $u(x)=1$ if $x>L_R$ and
$u(x)=u=1/g<1$, otherwise. The correlator of the $\theta$ field ordered
in imaginary time $T(x,y,\tau)\equiv <T_{\tau}{\theta(x,\tau)
\theta(y,0)}>$ can be shown to 
satisfy the following equation 
\begin{equation}
\{\frac{1}{v^2 u^2(x)} \partial^2_\tau + \partial^2_x \} T(x,y,\tau)=
-\frac{4 \pi}{v} \delta(x-y)\delta(\tau)
\label{4}
\end{equation}
under the boundary conditions $\partial_x T(x,y,\tau)|_{x=0}=0$ 
following from (\ref{1}). 
Fourier transform of this correlator $T(x,y,\omega)$
is symmetrical under $\omega\rightarrow -\omega$. It can be 
constructed from the solutions of the homogeneous equation corresponding
to Eq.(\ref{4})
\begin{eqnarray}
\{\frac{\omega^2}{v^2 u^2(x)} - \partial^2_x \} f_\omega(x)=
\{\frac{\omega^2}{v^2 u^2(x)} - \partial^2_x \} h_\omega(x)=0
\label{5}\\
T(x,y,\omega)=\frac{4 \pi}{v W(\omega)}
[\theta(x-y)f_\omega (x) h_\omega (y)+\theta(y-x)f_\omega (y) h_\omega (x)]
\nonumber 
\end{eqnarray}
if these solutions meet  boundary conditions:
$h'_\omega(0)=0,  f_\omega (x)=exp(-\omega x/v)$ at 
$x \rightarrow \infty$ and positive $\omega$. 
The Wronskian $W(\omega)$ is equal to $-f'_\omega (0) h_\omega(0)$
and, hence, $T(0,0,\omega)=-4 \pi /[v(ln f_\omega (0))']$. 
The only solution we need can be written as right going plus
reflected left going waves at $x<L_R$. The reflection amplitude 
$r \equiv r_{\theta}=-e^{-2 \eta},\  (\tanh(\eta)=1/u)$ for the $\theta$ field
is negative for the repulsive interaction. It
is related to the one $r_{\phi}$ for the $\phi$ field \cite{pon,safi}
as $r_{\theta}=
-r_{\phi}$ by the duality symmetry.
 Substituting this solution
one can find 
$T(0,0,\omega)=\frac{4 \pi u}{\omega} \tanh(\omega t_{R} + \eta)$ with
$t_{R}$ equal to the time of
travelling  from the junction to the right lead. 
Analytical continuation of this function $[-T(0,0,-i\omega+0)]$ is
the retarded Green function for the $\theta$ field. Imaginary 
part of
the latter multiplied by the Bose distribution function for holes
$1+f_B(\omega)$ and by a factor $(-2)$
coincides with the Fourier transform of the correlator at $\omega$. Then 
the particle density of states
(\ref{2}) in dimensionless units is obtained as
\begin{equation}
\rho_p(\varepsilon)=\frac{\rho_O}{2 \pi \gamma }
 \int^{+\infty}_{-\infty}
 d p exp \{i \varepsilon p + \int^\infty_\infty d \tilde{\omega}
e^{-\gamma |\tilde{\omega} |}
(1+f_B(\tilde{\omega}))\frac{e^{-i\tilde{\omega} p}-1}{\tilde{\omega}} 
\frac{Im \tan(\tilde{\omega}+i \eta)}{\tanh(\eta )}\}
\label{6}
\end{equation}
where the inverse temperature $\beta $ and energy $\epsilon, \omega $
were scaled as 
$\beta=1/(Tt_R), \ \varepsilon=t_{R} \epsilon, \ \tilde{\omega}=t_R \omega$ 
($T$ is the temperature in energy units). 
The dimensionless cut-off parameter $\gamma$ is
$(E_F t_{R})^{-1}$, and $p$ is dimensionless time $p=t/t_{R}$.
The hole density of states $\rho_h(\varepsilon)$ can be
found as $\rho_h(\varepsilon)=\rho_p(-\varepsilon)$.  

The correlator used in (\ref{6}) could be represented as a 
product after expansion of the imaginary part of the tangent
in a sum of exponents:
\begin{equation}
<\psi_r(0,p)\psi^+_{r}(0,0)>=\frac{E_F}{2 \pi v} 
\left(\frac{-i \gamma \pi /\beta}{\sinh((p-i\gamma)\pi/\beta} \right)^u
\prod _{n \ge 1}
\left(\frac{|\sinh((2n+i\gamma)\pi/\beta)|^2}
{\prod_{\pm}\sinh((2n\pm (p-i\gamma))\pi/\beta)} \right) ^{ur^n}
\label{7}
\end{equation}
This expression can be
easily understood as the product of
the contributions of the $2n$ length paths connecting $(0,p)$ and $(0,0)$
points and undergoing $n$ reflections 
from a $x=L_R$ non-elastic boundary
with the negative reflection amplitude $r=-e^{-2 \eta}$ 
and $n$ reflections 
from the $x=0$ elastic boundary with unit reflection amplitude.
The product has a good convergence due to exponential decrease
of $r^n$ with n. Substituting it into (\ref{6}) one can find 
low ($\varepsilon\ll 1$) and high ($\varepsilon \gg 1$) energy
behavior of the tunneling density of states. Similar calculations in the 
spinful case require
change of $Im[\tan(\tilde{\omega}+i \eta)]/\tanh(\eta )$ in (\ref{6}) into 
$(1+Im[\tan(\tilde{\omega}+i \eta)]/\tanh(\eta ))/2$. The product representation
of the correlator can
be obtained from (\ref{7}) if  
the exponent of the second multiplier $u$ is changed 
into $(u+1)/2$ and the exponents of all the rest of the product $ur^n$
into $ur^n/2$.

The calculation of the low energy expansion is tiresome but
straightforward. Up to the fourth order in energy it gives
in the spinless and spinful cases, respectively:
\begin{eqnarray}
\rho(\varepsilon)&=& c_1 \rho_O \gamma^{u-1}
\biggl(1+(1-g^2) \left[
\frac{\varepsilon^2}{2} + {1 \over 6}
\left( \frac{\pi}{\beta} \right)^2 \right]
+(1-g^2)(4.5 - 6.5 g^2)
\nonumber \\ 
&\times& \biggl[ 
0.104 \varepsilon^2 \left(\frac{\pi}{\beta} \right)^2
+  0.01 \varepsilon^4 \biggr] + 
(1-g^2)[1.34-1.9 g^2] \left({\pi \over \beta} \right)^4 \biggr)
\label{8}\\
\rho(\varepsilon)&=& \sqrt{c_1} \rho_O \gamma^{u-1}
\biggl(1+(1-g^2) \left[ {{\varepsilon^2} \over 4} + {1 \over 12}
\left( {\pi \over \beta} \right)^2 \right]
+(1-g^2)(2.22 - 3.22 g^2)
\nonumber \\
&\times& \left[ 
0.104 \varepsilon^2 \left( {\pi \over \beta} \right)^2 
+0.01 \varepsilon^4 \right] +
(1-g^2)[0.644-0.926 g^2]\left({\pi \over \beta} \right)^4 \biggr)
\label{10}
\end{eqnarray}
where the constant $c_1=\prod_{n>0}(2n)^{2ur^n}$ modifies a renormalization
parameter $\gamma^{u-1}$ and is model dependent.
Except for this parameter the functions $\rho(\varepsilon)$ and, hence,
all coefficients of (\ref{8},\ref{10}) in the curly brackets are universal
functions of $g$ and $t_{R}$. The coefficients in the spinful case
are approximately two times less than the ones in the spinless. 
In the second order this correspondence is exact.

The high energy behavior of the tunneling density at zero temperature
and positive $\varepsilon$
can be obtained making transform of the counter of integration in (\ref{6})
into a sum of the contours going around the cuts of the correlator (\ref{7})
in the complex time plane
from $i \infty+2n$ to $2n $ and back. The result takes the form:
$\rho(\varepsilon)/\rho_O=\frac{\gamma^{u-1}}{\pi}( \sin(\pi u)
\Gamma(1-u) \varepsilon^{u-1} + 2 r(\varepsilon))$ in the spinless case. 
The spinful expression just needs replacement of $u$ by $(u+1)/2$ 
in the spinless one. It reveals that 
finite length corrections 
\begin{eqnarray}
r(\varepsilon)&=&
\sum_{n>0}a_n \sin(\pi u r^n) \Gamma(1-ur^n)
\cos(2n \varepsilon - {\pi \over 2}(1-r^n))
\varepsilon^{ur^n-1}  \mbox{ \ \ (spinless)}
\label{9}\\
r(\varepsilon)&=&
\sum_{n>0} b_n \sin(\pi u r^n/2) \Gamma(1-ur^n/2)
\cos(2n \varepsilon - {\pi \over 4}(2-r^n))
\varepsilon^{ur^n/2-1}  \mbox{\ \ (spinful)}
\label{11}
\end{eqnarray}
to the scale-invariant power law dependence display an interference 
structure. Here the constant $a_n$ in the spinless 
density of states (\ref{9}) is given by  $ a_n ^{-1}=2^u n^{u(1-r^n)}
\prod_{m \neq n >0} \left| 1-\left({n \over m} \right)^2 
\right| ^{ur^m}$ 
and the constant $b_n$ in the spinful density (\ref{11}) is
$b_n=\sqrt{a_n/(2n)}$. Both of them 
are on the order of 1 at small $n$ and decrease
with increase of the number. However, a quick convergence in (\ref{9}, \ref{11})
is mostly due to the $u r^n$ factor. The first two terms of the sums 
bring the leading contribution.
They describe weakly decaying oscillations with the periods 
equal to $2t_{R} \;(n=1)$ and $4t_{R} \;(n=2)$, respectively. 
The second term
always dominates at large energies as $r<0$. However, the crossover energy
is exponentially large at small $u$. 
Comparing coefficients of the first two oscillating terms of the sum
in (\ref{11}) one can gather that the finite length oscillations of the
$2t_{R}$ period dominate the ones of the $4t_{R}$ period above 
$\varepsilon \approx 1$ unless $u>1.8$ in the spinless case or $u>2.1$
in the spinful one.
It is in accordance with the numerical calculations \cite{cond}.
This structure in the density of states may be understood
as a quantization of the
plasmonic modes  inside the wire \cite{naz}.

{\it Conductance} - It is convenient below to redefine the dimensionless 
energy in units of
the inverse total traversal time as $T \equiv t_0 T,\ V \equiv t_0 V$. 
Low energy conductance is an analytical function
of $T,V < \zeta^{-1}, (1-\zeta)^{-1}$. Integrating the low energy tunneling 
density of states 
(\ref{8},\ref{10}), it can be found
in the spinless and spinful cases, respectively:
\begin{eqnarray}
\sigma=\frac{ c_1^2 \tilde{\gamma}^{2(u-1)}}{R_O  \zeta_1^{u-1}}
\biggl(&1&+(1-g^2) \zeta_2 \left[{V^2 \over 2} +
{(\pi T)^2 \over 3} \right]
+\biggl[ {{(1-g^2)^2}\over 6} \zeta_1^2 
+(1-g^2){{9-13 g^2} \over {16}} \zeta_3 \biggr] (\pi  T V)^2 \biggr)
\nonumber \\
+\Delta_T \sigma
\label{12'}\\
\sigma=\frac{ c_1 \tilde{\gamma}^{u-1}}{R_O  \zeta_1^{(u-1)/2}}
\biggl(&1&+{{1-g^2}\over 2} \zeta_2 \left[{V^2 \over 2}+
{(\pi T)^2 \over 3} \right]
+\biggl[ {{(1-g^2)^2}\over 4} \left( {\zeta_1^2 \over 6}
- {\zeta_3 \over {64}}\right) 
\nonumber \\
&+&  (1-g^2){{9-13 g^2} \over {32}} \zeta_3 \biggr] (\pi  T V)^2 \biggr)
+\Delta_T \sigma
\label{12}
\end{eqnarray}
where $\tilde{\gamma}^{-1}=t_0 E_F$ and $\zeta_1=\zeta (1-\zeta), \zeta_2=1-2 \zeta_1, 
\zeta_3=\zeta^4+(1-\zeta)^4 $ are geometrical coefficients determined 
by the only parameter $\zeta$ of the barrier position and 
$R_O^{-1}$ is a free electron conductance of the junction.
$\Delta_T \sigma$ is produced by the finite
difference between the temperatures of the right and left reservoirs
$\Delta T=T_R-T_L$ and $T=(T_R+T_L)/2$. In the lowest linear order it is 
non-zero unless $\zeta=1/2$ and reads for the spinless and spinful conductance,
respectively:
\begin{eqnarray}
\Delta_T \sigma&=&\frac{c_1^2 \tilde{\gamma}^{2(u-1)}}{R_O  \zeta_1^{u-1}}
{{(1-g^2)(9-13 g^2)}\over 72} \zeta_4  V^2 \pi^2 T \Delta T
\label{13'}\\
\Delta_T \sigma&=&\frac{c_1 \tilde{\gamma}^{u-1}}{R_O  \zeta_1^{(u-1)/2}}
{{1-g^2}\over 12} 
(0.74 - 1.073g^2)\zeta_4  V^2 \pi^2 T \Delta T
\label{13}
\end{eqnarray}
where $\zeta_4=\zeta^4-(1-\zeta)^4$. For $\zeta =1/2$ $\Delta_T \sigma$
in(\ref{13'}), (\ref{13}) vanishes and the conductance is
always increasing  due to $(\Delta T)^2$ term 
as the temperature difference is introduced. 
Relation of the $T^2$ term coefficient to the $V^2$ one 
is universal. Following Weiss's consideration \cite{weiss} one can prove  
it is equal to $(2 \pi^2)/3$ in all orders of the perturbation expansion in the 
tunneling amplitude \cite{prep}. This is
similar to Wilson relation in the Kondo problem. One can see from 
(\ref{12'}), (\ref{12}) that
both the interaction constant $g$ and the traversal time $t_0$ may be
determined from the low energy expansion by
comparing coefficients for the terms of the second and fourth order in energy.

Next we consider the high energy behavior of the conductance.
To describe the latter for $ V \gg 1/(1-\zeta), 1/\zeta $ it is convenient
to start from a 
representation of the current in the form: $<J>=|A|^2 \int d t 
(e^{i V t}-e^{-i V t}) <\psi_R(0,t) \psi^+_R(0,0)>
<\psi^+_L(0,t) \psi_L(0,0)>$. The correlators may be expressed in the form
(\ref{7}). Making asymptotic integration we come to the conductance
which in the spinless case could be written as $\partial J/ \partial V=
(\pi R_O)^{-1}(t_0 E_F)^{2(1-u)}[\sin{\pi(2 u-1)}\Gamma(2-2u)V^{2u-2}+\Delta \sigma(V)]$.
Expression for the spinful conductance is obtained by
replacing $u$ by $(u+1)/2$.
At zero temperature the finite length corrections $\Delta \sigma(V)$ of spinless 
and spinful conductance
may be found for irrational values of $\zeta$, respectively, as
\begin{eqnarray}
\Delta \sigma(V)=2V^{2u-2} \sum_{n>0} 
\biggl[\frac{\cos(2nV \zeta + 2\varphi_{n,R})}{(V \zeta)^{2u(1-r^n/2)-1}
g^2_n\left({\zeta \over {1-\zeta}}\right)} &+& 
\frac{\cos(2nV (1-\zeta) + 2\varphi_{n,L})}{(V (1-\zeta))^{2u(1-r^n/2)-1}
g^2_n\left({{1-\zeta} \over \zeta}\right)} \biggr] \nonumber \\
&\times& a_n
\sin(\pi u r^n) \Gamma(1-ur^n)
\label{14'}\\
\Delta \sigma(V)=2V^{u-1} \sum_{n>0} 
\biggl[\frac{\sin(2nV \zeta + \varphi_{n,R})}{(V \zeta)^{u(1-r^n/2)}
g_n\left({\zeta \over {1-\zeta}}\right)} &+& 
\frac{\sin(2nV (1-\zeta) + \varphi_{n,L})}{(V (1-\zeta))^{u(1-r^n/2)}
g_n\left({{1-\zeta} \over \zeta}\right)} \biggr] \nonumber \\
&\times& b_n
\sin(\pi u r^n/2) \Gamma(1-ur^n/2)
\label{14}
\end{eqnarray}
where $g_n^2(x)=(2n)^{u-1} \prod_{m >0} 
\left| 1-\left(xn/m \right)^2 \right|^{ur^m}$.
The phase shifts are $(r^n-(u-1)r^{[nx]})\pi/4$ ($[nx]$ denotes the integer
part of $nx$)
 with $x=\zeta/(1-\zeta)$ for
$\varphi_{n,R}$ and $x=(1-\zeta)/\zeta$ for $\varphi_{n,L}$. 
Again one can see that only the $n=2$ term is important at high
$V$. However,  if $-r \ll 1$ the $n=1$ oscillations 
dominate over a large range 
of energy above $1$. With decrease of $\zeta < 1/2$  the oscillations
of the $\pi/(\zeta t_0)$ periodicity acqure much larger amplitude than those 
of $\pi/(1-\zeta)$. 
 
If $\zeta$ is rational: $\zeta/(1-\zeta)=n_2/n_1$ the resonant enhancement 
of the oscillations with the frequency $2n_1 \zeta V = 2 n_2 (1-\zeta)V$
occurs. Since only lowest $n$'s  contribute to the sum (\ref{14})
this resonance  is important when both $n_i$'s are small: $\zeta=1/2,1/3,2/3$.
For $\zeta=1/2$ the finite length correction of the
spinless and spinful conductances reads
\begin{eqnarray}
\Delta\sigma(V)&=&V^{2u-2} \sum_{n>0} 
4 n a^2_n \cos(2nV\zeta + \pi ( r^n-1))
\sin(2\pi u r^n) \Gamma(1-2ur^n)(V \zeta)^{2u(r^n-1)-1}
\label{15'}\\
\Delta\sigma(V)&=&V^{u-1} \sum_{n>0} 
2 a_n \sin(2nV\zeta + \pi r^n/2)
\sin(\pi u r^n) \Gamma(1-ur^n)(V \zeta)^{u(r^n-1)}
\label{15}
\end{eqnarray}
The resonant enhancement strengthens oscillations of the $4\zeta$ frequency
and weakens the ones of the $2\zeta$. 
These interference structures are shown in Fig.1 for spinful electrons.

With decrease of $V$ the high voltage behavior of the conductance 
(\ref{14'}), (\ref{14})
will meet the low voltage one (\ref{12'}), (\ref{12}). However, 
if $\zeta \ll 1/2$ there is a region $1/(1-\zeta) \ll V \ll 1/\zeta $ 
where conductance relates to the low and high energy tunneling 
densities of states of the right and left parts of the wire, appropriately.
In the leading order it is described by 
$\sigma=\rho_R(0) \rho((1-\zeta) V)/(R_O \rho_O^2)$.

In summary, we considered suppression of the 
conductance through a tunneling barrier in a 1D channel constriction 
in the ITLL model accounting for the finite length of the wire.
Starting from the zero energy the conductance increases analytically
as $T^2, V^2$ and higher even integer degrees. The relations between 
all coefficients  
are determined by the constant of interaction and two traversal times.  
At voltages higher than inverse of the 
sum of the traversal times conductance approaches
the infinite length scale-invariant dependence. The deviation from the 
latter produced by the finite length contribution 
decays as a negative degree of the voltage oscillating with the 
periodicities related to double traversal times for a weak repulsion
and to quadrupole ones if the interaction become stronger.

We acknowledge S. Tarucha for useful discussions.
It is a special pleasure for one of us (V.P.) 
to thank T. Iitaka for his help in conducting
calculations.
This work was supported by the Center of Excellence  and 
partially by the fund of the JSPS for development of collaboration
between the former Soviet Union and Japan.

\figure{Dependences of the finite length correction 
$(E_F t_0)^{(1-u)} \Delta \sigma(V)/R_O$ to the differential conductance
on volage $V$ measured in $\pi$ over traversal time $t_0$ unit 
for spinful electrons at $u=1.396$: the solid line relates to the symmetrical
position of the weak link; the dotted line to the case when the ratio between
the lengths of the right and left shoulders is 1/4.}

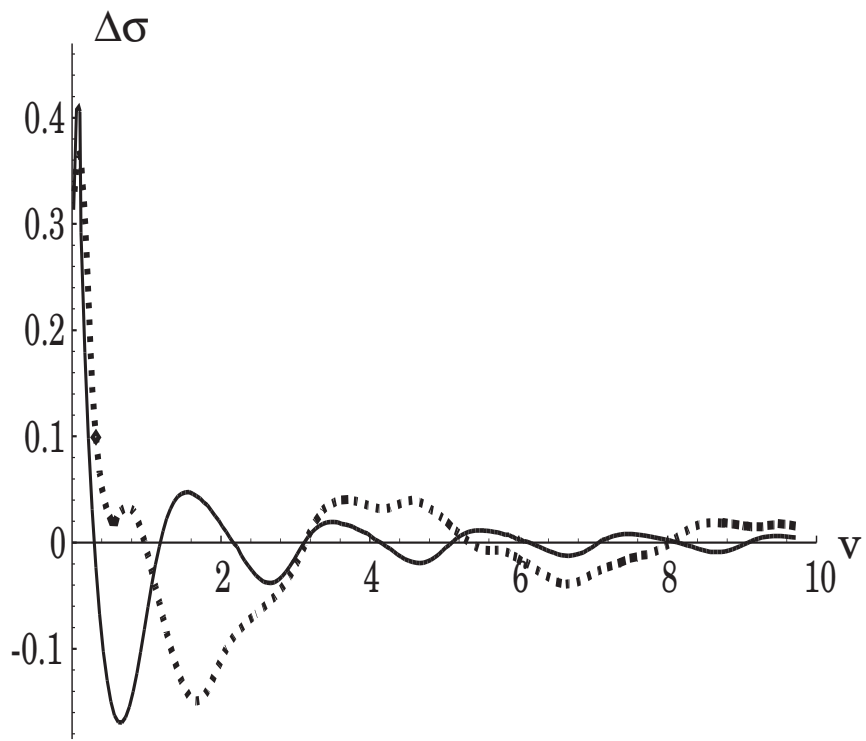
\end{document}